\documentclass[preprint]{article}
\usepackage{subcaption}
\usepackage{hyperref}
\usepackage{amsmath,graphicx,mlspconf}
\usepackage{amsfonts}
\usepackage{url}
\usepackage{xcolor}
\usepackage{booktabs}
\usepackage{color, colortbl}
\newcommand{\ci}[1]{\tiny{±#1}}
\definecolor{Gray}{gray}{0.94}
\definecolor{color1}{HTML}{FFF8EF}
\definecolor{color2}{HTML}{FFD8AD}
\definecolor{color3}{HTML}{F7E9DB}
\usepackage{amsthm}
\usepackage{mathtools}
\usepackage{thmtools}
\usepackage[mathscr]{eucal}
\copyrightnotice{979-8-3503-2411-2/25/\$31.00 {\copyright}2025 IEEE}

\toappear{2025 IEEE International Workshop on Machine Learning for Signal Processing, Aug.\ 31-- Sep.\ 3, 2025, Istanbul, Turkey}

\title{AudioMAE++: learning better masked audio representations with SwiGLU FFNs}
\name{%
   Sarthak Yadav$^{\star \ast}$
   \qquad Sergios Theodoridis$^{\star \dagger}$
   \qquad Zheng-Hua Tan$^{\star \ast}$%\thanks{Thanks to XYZ agency for funding.}
}
\address{
   $^{\star}$ Aalborg University, Denmark \quad $^{\ast}$ Pioneer Centre for AI, Denmark \\ $^{\dagger}$ National and Kapodistrian University of Athens, Greece
}

\begin{document}
% \ninept

\maketitle

\begin{abstract}
Masked Autoencoders (MAEs) trained on audio spectrogram patches have emerged as a prominent approach for learning self-supervised audio representations. While several recent papers have evaluated key aspects of training MAEs on audio data, the majority of these approaches still leverage vanilla transformer building blocks, whereas the transformer community has seen steady integration of newer architectural advancements. In this work, we propose AudioMAE++, a revamped audio masked autoencoder with two such enhancements, namely macaron-style transformer blocks with gated linear units. When pretrained on the AudioSet dataset, the proposed AudioMAE++ models outperform existing MAE based approaches on 10 diverse downstream tasks, demonstrating excellent performance on audio classification and speech-based benchmarks. 
The proposed AudioMAE++ models also demonstrate excellent scaling charecteristics, outperforming directly comparable standard MAE baselines with up to $4\times$ more parameters.
% Nulla ex metus, imperdiet nec mauris at, accumsan egestas sapien. Nullam maximus porttitor tristique. Maecenas id odio eget magna vehicula iaculis. Nullam pretium mollis turpis in lacinia. Nam laoreet sit amet ante eu blandit. Quisque dolor erat, scelerisque nec lobortis vel, varius eu metus.
\end{abstract}

\section{Introduction}
\label{sec:intro}

In recent years, self-supervised learning (SSL) has garnered significant interest for learning deep feature representations by leveraging unlabeled data. The widespread adoption of SSL is led by masked predictive modeling approaches, where the objective is to correctly predict masked portions of the input and/or latent representations using the unmasked subset. Several groundbreaking works leveraging masked predictive modeling across text \cite{bert2019}, vision \cite{xie2022simmim, bao2022beit}, speech and audio \cite{baevski2020wav2vec, chen2022wavlm, hsu2021hubert}, as well as work on unified frameworks across modalities \cite{baevski2023efficient} have been proposed.

The Masked Autoencoder (MAE) \cite{he2022masked} is one such recent approach. MAEs were recently proposed for learning visual representations from masked non-overlapping image patches linearly projected to a fixed dimension. 
In contrast to existing visual representation learning approaches of the time, MAEs differed in three key aspects: (i) \textit{High masking rates:} to counter the high spatial correlation between image pixels and to minimize extrapolation from redundant neighbouring patches, MAEs adopt a very high masking ratio, masking out over 75\% of the input patches; (ii) \textit{Encoding only the visible patches}: as opposed to existing approaches that encoded both visible and masked patches (filled in-place with a learnable mask token), MAE encoders generate contextual representations based only on the visible patches. 
Learnable mask tokens are then filled in to restore the original patch order, and a transformer based decoder then attempts to reconstruct the masked patches; (iii) \textit{Asymmetrical architecture:} given the reduced encoding complexity, asymmetrically large encoders can be paired with smaller decoders, and the decoders are discarded after pretraining. 

Given these inherent advantages, MAEs have seen wide-spread adoption across several domains, including vision \cite{he2022masked, feichtenhofer2022masked, tong2022videomae, woo2023convnext}, remote sensing \cite{scalemae}, depth estimation \cite{bachmann2022multimae} and medical imaging \cite{zhou2023self}, to name a few. The audio domain has not remained untouched, with several recent works leveraging a masked autoencoder framework for learning general-purpose self-supervised audio representations. 
These works investigate key aspects of modeling audio data \cite{niizumi2022masked}, explore joint discriminative and generative objectives \cite{baade_mae-ast_2022}, the impact of local-global attention \cite{huang2022masked, yadav2024masked}, as well as scaling out to large datasets and model sizes \cite{dinkel24b_interspeech}. 
However, while the transformer community has moved forward with newer advancements, such as gated linear units \cite{shazeer2020glu}, newer length-extrapolatable positional embeddings \cite{sun2023length, su2024roformer} and macaron-style feedforward networks (FFNs) \cite{lu2020understanding}, majority of such existing MAE approaches for audio still leverage standard transformer building blocks, namely fixed positional embeddings and standard point-wise feedforward modules. Thus, a formal evaluation of these advances, and whether they offer any improvements in masked modeling of audio representations, is pending.
\begin{figure}
    \centering
    \includegraphics[trim={2em 2em 2.5em 4em}, clip, width=0.55\linewidth]{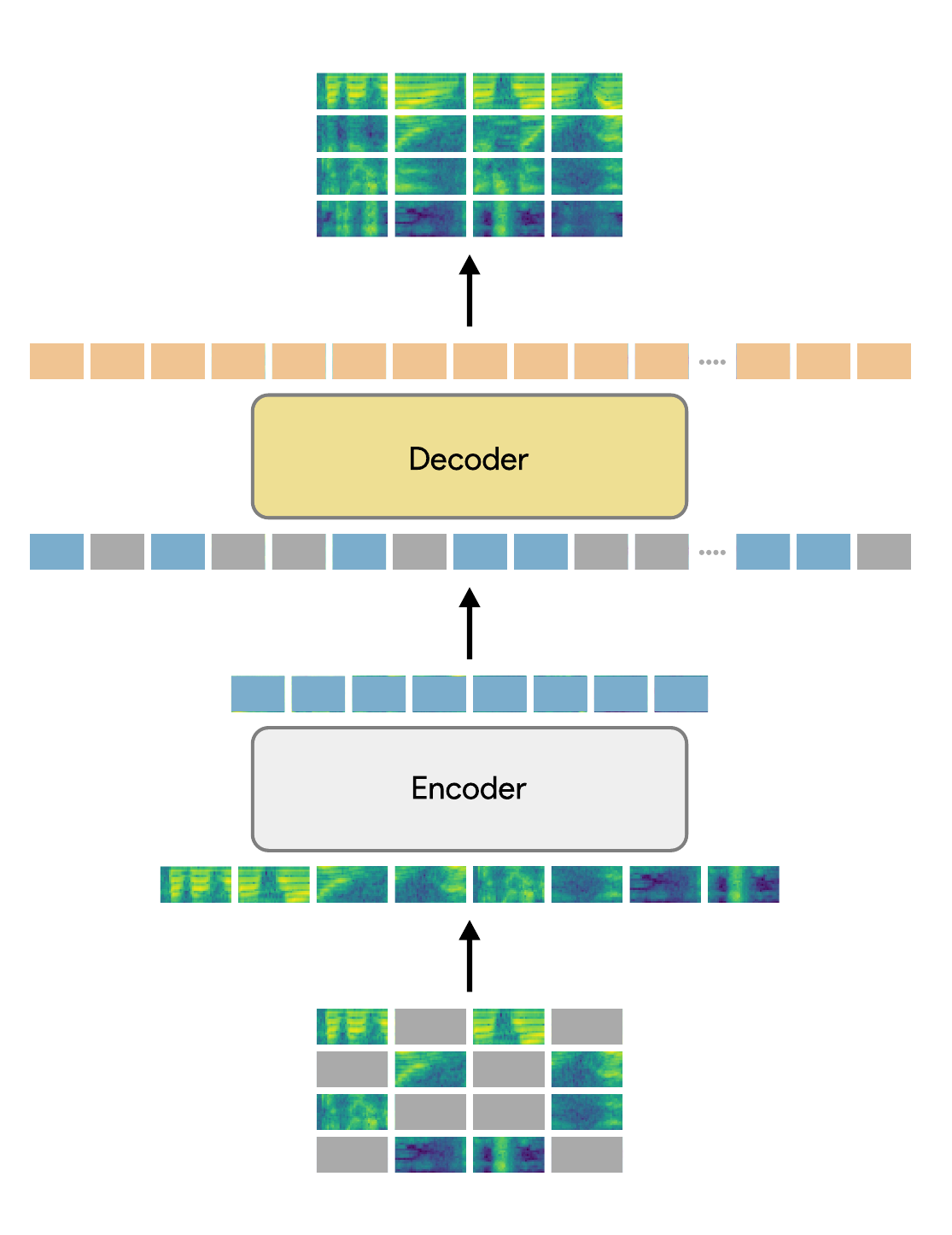}
    \caption{Outline of a masked autoencoder.}
    \label{fig:maeoutline}
\end{figure}
In this work, we propose AudioMAE++, a revamped masked autoencoder for learning general-purpose audio representations. In AudioMAE++, we investigate the impact of some of the above-mentioned advancements, namely SwiGLU FFNs and rotary positional embeddings, which have previously shown promise for convolution-free automatic speech recognition (ASR) \cite{hou24_interspeech}. Pretrained on the AudioSet dataset and  evaluated on 10 diverse downstream audio classification tasks, the proposed AudioMAE++ approach consistently achieves better performance and offers better scalability, outperforming up to $4\times$ larger standard MAE baselines. Code and pretrained models can be found \href{https://github.com/SarthakYadav/audiomae-plusplus-official}{here}.

\section{Approach}
\label{sec:approach}
\subsection{Prerequisites}
\label{ssec:prereq}

Before going forward, we briefly describe the three architectural changes incorporated into the proposed AudioMAE++.
\textbf{(i) Rotary Positional Embeddings:} 
Given that the transformer architecture and the underlying self-attention module capture position independent contextual representations, infusing positional information appropriately in the transformer is key for modeling sequential data. 
% Several approaches exist for incorporating positional embeddings into transformers, such as fixed sinusoidal positional embeddings \cite{vaswani2017attention} and relative positional encodings \cite{shaw-etal-2018-self, transformerxl}. 
Several approaches exist for incorporating positional embeddings into transformers, such as fixed sinusoidal positional embeddings \cite{vaswani2017attention} and relative positional encodings \cite{transformerxl}. 
However, these approaches have limited length extrapolation capabilities and often require explicit extrapolation strategies when used on input sequence lengths unseen during training. 
Several positional embedding approaches have been investigated to address these issues, amongst them being the rotary positional embeddings (RoPE) \cite{su2024roformer}. 
RoPE uses a rotation matrix to encode absolute position into the token embedding, while simultaneously embedding relative positional information, allowing better sequence-length flexibility.\newline
% \textbf{(ii) Macaron-style FFN:} In contrast to the standard transformer block which consists of a single FFN in each block, a macaron-style \cite{lu2020understanding} block sandwiches the self-attention module between two FFN blocks and has shown to exhibit better language \cite{lu2020understanding} and speech \cite{gulati20_interspeech, hou24_interspeech} modeling capabilities.\newline
\textbf{(ii) Macaron-style FFN:} In contrast to the standard transformer block which consists of a single FFN in each block, a macaron-style \cite{lu2020understanding} block sandwiches the self-attention module between two FFN blocks and has shown to exhibit better language \cite{lu2020understanding} and speech \cite{hou24_interspeech} modeling capabilities.\newline
\textbf{(iii) Gated Linear Units (GLU):} Gated linear units (GLU) \cite{dauphin2017language} are a neural network layer where the output is an element wise product of two linear projections of the same input, with one of the projections possessing a sigmoid activation. Several variants of GLUs exist, including the SwiGLU, where the sigmoid activation is replaced by the Swish activation function \cite{ramachandran2017searching} ($F_{\text{SwiGLU}}(x; W, V, O) = (\text{Swish}(xW) \otimes xV)O$), and they have been shown to improve performance in sequence-to-sequence transformers \cite{shazeer2020glu, hou24_interspeech}.

\subsection{AudioMAE++}
\label{ssec:audiomaepp}
\begin{figure}
    \centering
    \includegraphics[trim={2.5em 3.0em 3.5em 3em}, clip, width=0.7\linewidth]{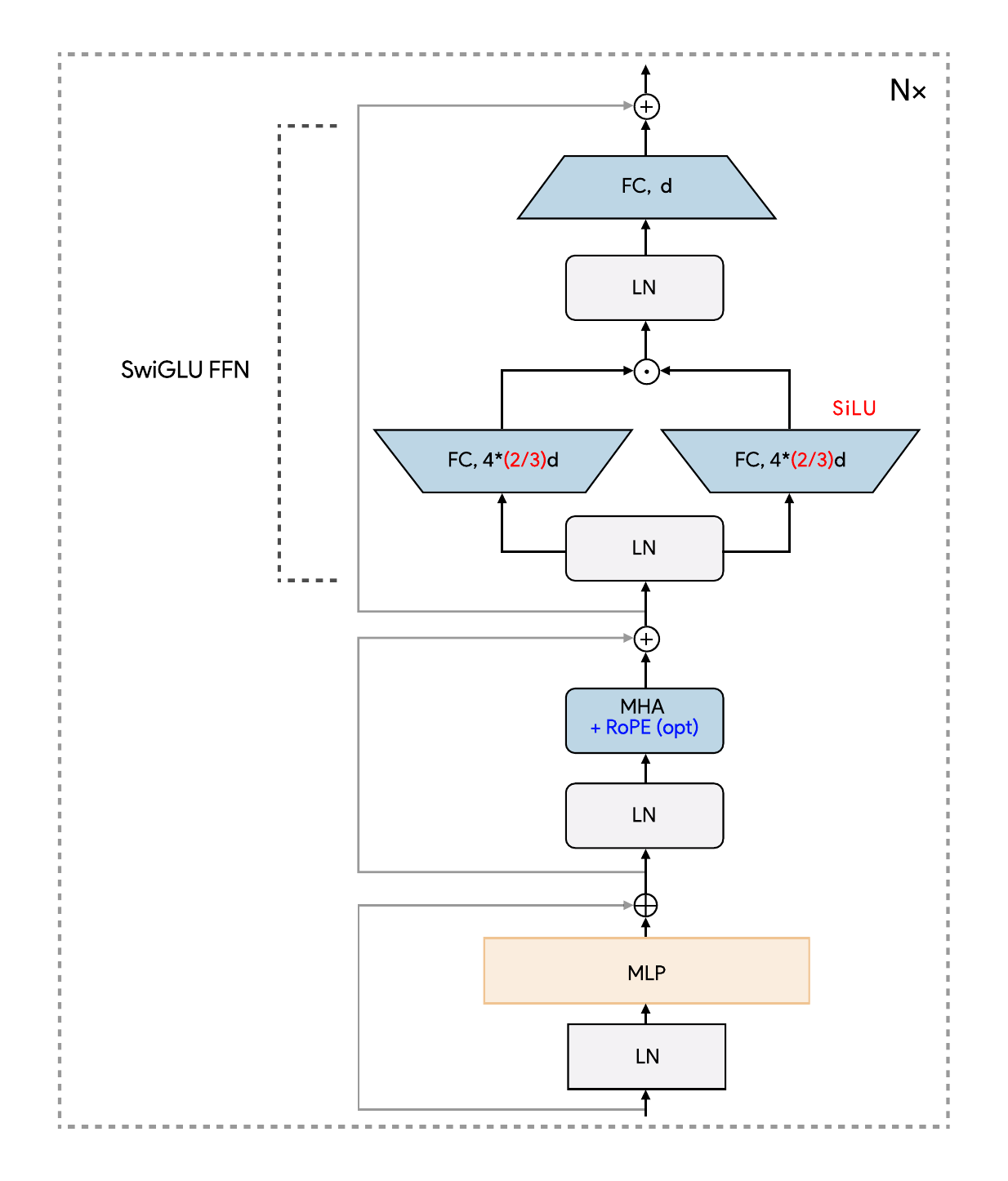}
    \caption{The transformer++ building block. LN, MHA and FC stand for LayerNorm, Multihead Attention and Fully Connected layer, respectively.}
    \label{fig:transformerpp}
\end{figure}
Figures~\ref{fig:maeoutline} and \ref{fig:transformerpp} show an overview of the proposed AudioMAE++ model, outfitted with transformer++ \cite{hou24_interspeech} based blocks with optional RoPE embeddings.
\newline\textbf{Patch creation and random masking:} For an input spectrogram $x \in \mathbb{R}^{\mathtt{T} \times \mathtt{F}}$, where the axes denote time and frequency, respectively, a patch embedding layer produces $N$ non-overlapping patches of shape $t\times f$, and projects them to $d_m$-dim space, resulting in $x_p \in \mathbb{R}^{N \times d_m}$. 
A standard practice is to add fixed sinusoidal 2-d positional embeddings to the patches at this stage \cite{dosovitskiy2021an, he2022masked}. In the proposed AudioMAE++ module, we experiment with (optional) per-block rotary positional embeddings instead. A learnable \textit{cls} token is then prefixed to the input patches, which is in-line with existing works \cite{he2022masked,niizumi2022masked,huang2022masked} as well as facilitates easy fine-tuning. 
Similar to \cite{yadav2024masked}, 80\% of the patches are then randomly masked and dropped,  resulting in $x_p' \in \mathbb{R}^{0.2N \times d_m}$ patches.
\newline\textbf{Encoder:} The encoder, which is a stack of transformer++ \cite{hou24_interspeech} blocks, ingests the partial input patches and generates encoded representations $z_{enc} = enc(x_p')$. 
In contrast to the standard transformer block, the proposed block sandwiches the multi-head attention (MHA) block between a single-hidden layer MLP and a SwiGLU FFN in a macaron-style layout \cite{lu2020understanding} (shown in Fig.~\ref{fig:transformerpp} and Eq~\ref{eq:ffn} for a given input $x_i$), and thus has more parameters.
\begin{align}
    \label{eq:ffn}
    x_i' = x_i + \frac{1}{2}\text{MLP}(\text{LN}(x_i)),\\
    x_i' = x_i' + \text{MHA}(\text{LN}(x_i')),\\
    y_i' = \text{LN}(x_i' + \frac{1}{2}F_{\text{SwiGLU}}(x_i'))
\end{align}
\newline\textbf{Decoder:} The decoder accepts the contextualized representations $z_{enc}$ and fills in a learnable masked token at the correct indices to restore the correct patch order, before passing them through a stack of transformer++ blocks with dimensions $d_{dec}$ and a final $(t.f)$-dimensional linear projection. The first index corresponding to the cls token is then discarded to obtain the output $y \in \mathbb{R}^{N \times t.f}$. 
The mean-squared error (MSE) between the input spectrogram and the corresponding reconstructions corresponding to the masked positions is used for pretraining.

\section{Experiments}
\label{sec:exps}

\subsection{Datasets used and aggregated performance metric}
\textbf{Pretraining:} We use the AudioSet dataset (AS) \cite{gemmeke2017audio} for pretraining our models. With about 5000 hours of audio data, AudioSet comprises of roughly 2 million 10-second long weakly-labeled YouTube clips spanning 527 classes.
\newline\textbf{Downstream Evaluation:} For evaluating the proposed approach on a wide variety of downstream tasks, we utilize a 10-task subset of the HEAR benchmark\footnote{\url{https://hearbenchmark.com}}, as shown in Table~\ref{tab:tasks}. This selection of tasks is inline with several recent works \cite{yadav2024masked, yadav2024audiomambaselectivestate} and spans music, pitch perception, speech and audio classification domains. 
% In line with several Following recent work \cite{yadav2024masked, yadav2024audiomambaselectivestate}, we use the following subset of tasks proposed as part of the HEAR benchmark \cite{turian_hear_2022} for downstream evaluation: Beijing Opera \cite{beijingopera, turian_hear_2022}, Crema-D \cite{cao2014crema}, ESC-50 \cite{esc50}, LibriCount \cite{stoter2018classification}, Mridangam Stroke and Tonic \cite{mridangamds}, NSynth Pitch 5h \cite{turian_hear_2022, nsynth2017}, Speech Commands 5h \cite{turian_hear_2022, warden2018speech}, FSD50K \cite{fonseca2021fsd50k} and VoxLingua107 \cite{kim2018vocal}. provides a brief overview of these tasks.
\begin{table}[h]
  \setlength\tabcolsep{1pt}
  \centering
  \scriptsize
  \begin{tabular}{lcccc}
    \toprule
    % \multicolumn{2}{c}{Part}                   \\
    % \cmidrule(r){1-2}
    ID & Name     & Description   & \#Classes & \#Hours  \\
    \midrule
    BO & Beijing Opera  & percussion instrument classification & 4 & 0.3 \\
    CD & Crema-D    & emotion recognition & 6 & 10 \\
    E50 & ESC-50  & environmental sound classification & 50 & 2.77 \\
    LC & LibriCount & counting speakers (classification) & 10 & 8  \\
    Mri-S & Mridangam Stroke & classifying Mridangam \textit{strokes} & 10 & 1.57  \\
    Mri-T & Mridangam Tonic  & classifying Mridangam \textit{tonics} & 6 & 1.57  \\
    NS-5h & NSynth Pitch 5h  & pitch classification & 88 & 5.5  \\
    SC-5h & SpeechCommands 5h  & keyword spotting & 12 & 6.5  \\
    F50K & FSD50K & multilabel audio tagging & 200 & 100  \\
    VL & VoxLingua107 Top10 & spoken language identification & 10 & 5  \\
    \bottomrule
  \end{tabular}
  \caption{Evaluated downstream tasks.}
  \label{tab:tasks}
\end{table}
\newline\textbf{Aggregated Performance Metric:} Given the wide variety of downstream tasks with different individual performance metrics and scales, as well as the wide variety of models evaluated, we decided to use the aggregated normalized score as proposed by \cite{yadav2024masked}. For a model $m$, overall score $s(m) \in [0,100]$ is given as: $s(m) = \frac{1}{|T|}\sum_{t \in T} \frac{x_t(m) - min_t}{max_t - min_t} * 100$, where $x_t(m)$ denotes performance of the model $m$ on task $t$, and $min_t$ and $max_t$ represent the worst and the best performance across all models on the task, thus taking into account the relative performance amongst all evaluated representations.
\begin{table*}[t]
    \vspace{-1em}
    \setlength\tabcolsep{2.pt}
    \small
    \centering
    \begin{tabular}{lcccccccccccccc}
    \toprule
    \multicolumn{3}{l}{} &\multicolumn{4}{c}{Music \& Pitch}  &\multicolumn{4}{c}{Speech-based tasks} &\multicolumn{2}{c}{Audio}\\
    \cmidrule(lr){4-7} \cmidrule(lr){8-11} \cmidrule(lr){12-13}
    \multicolumn{1}{l}{Model} & \multicolumn{1}{l}{Data} & \#M Params & BO & Mri-S & Mri-T & NS-5h & CD & LC & SC-5h & VL & E50 & F50K & $s(m)$ \\
    \midrule
    \multicolumn{3}{l}{\textbf{Supervised Baselines}} & \multicolumn{5}{l}{}\\
    \rowcolor{Gray}
    HEAR-Naive & - & - & 52.6\ci{2.4} & 38.0\ci{1.3} & 36.4\ci{1.9} & 18.6\ci{4.4} & 30.9\ci{0.8} & 33.5\ci{1.1} & 8.5\ci{0.4} & 11.2\ci{0.5} & 5.8\ci{0.2} & 7.1\ci{0.2} & 5.0\ci{0.7}\\
    % \midrule
    % \multicolumn{3}{l}{\textbf{Supervised}} & \multicolumn{5}{l}{}\\
    PaSST-Base \cite{koutini22_interspeech} & AS & $86$ & 94.9\ci{0.5} & 96.5\ci{0.1} & 87.6\ci{0.6} & 23.3\ci{0.9} & 61.0\ci{0.3} & 60.1\ci{0.2} & 66.6\ci{1.4} & 25.5\ci{0.8} & \textbf{94.8\ci{0.3}} & \textbf{64.2\ci{0.1}} & 73.5\ci{0.4}\\

    \midrule
    \multicolumn{3}{l}{\textbf{SSL}} & \multicolumn{5}{l}{}\\
    \rowcolor{Gray}
    W2V2-large \cite{baevski2020wav2vec} & VP & $315.4$ & 93.1\ci{0.7} & 93.9\ci{0.1} & 77.4\ci{0.2} & 42.0\ci{1.0} & 66.9\ci{0.4} & 62.4\ci{0.3} & 87.6\ci{0.5} & 53.6\ci{1.0} & 60.1\ci{0.5} & 34.2\ci{0.1} & 74.0\ci{0.4}\\
    WavLM-large \cite{chen2022wavlm} & Mix & $315.4$ & 96.4\ci{0.5} & 96.8\ci{0.1} & 89.5\ci{0.1} & 53.7\ci{0.5} & 57.2\ci{0.2} & 61.1\ci{0.3} & 46.2\ci{0.8} & 23.7\ci{0.9} & 47.9\ci{0.4} & 29.0\ci{0.1} & 64.1\ci{0.2}\\
    \rowcolor{Gray}
    HuBERT-large \cite{hsu2021hubert} & LL & $315.4$ & 94.1\ci{0.7} & 95.3\ci{0.1} & 83.5\ci{0.3} & 19.3\ci{0.8} & 70.7\ci{0.1} & 59.9\ci{0.2} & 83.2\ci{0.7} & \textbf{66.1\ci{0.9}} & 60.3\ci{0.4} & 31.5\ci{0.1} & 73.4\ci{0.3}\\
    BEATs-iter3 \cite{chenbeats23} & AS & $90.0$ & 94.0\ci{0.8} & 94.7\ci{0.1} & 95.8\ci{0.1} & 69.4\ci{0.8} & 67.3\ci{0.2} & 68.0\ci{0.2} & 85.2\ci{0.3} & 38.5\ci{1.0} & 83.7\ci{0.3} & 53.6\ci{0.2} & 85.7\ci{0.3}\\
    \rowcolor{Gray}
    SSAST \cite{gong2022ssast} & Mix & $89.0$ & 93.4\ci{0.9} & 96.7\ci{0.1} & 96.3\ci{0.1} & 66.8\ci{0.7} & 56.5\ci{0.2} & 60.7\ci{0.3} & 53.5\ci{1.3} & 28.5\ci{0.9} & 68.4\ci{0.4} & 38.2\ci{0.1} & 71.7\ci{0.2}\\
    SSAM-Tiny \cite{yadav2024audiomambaselectivestate} & AS & $4.8$ & 93.7\ci{0.8} & 97.1\ci{0.1} & 94.9\ci{0.1} & 62.0\ci{0.7} & 61.8\ci{0.3} & 59.2\ci{0.4} & 74.8\ci{0.4} & 27.8\ci{1.0} & 70.6\ci{0.2} & 41.3\ci{0.2} & 74.8\ci{0.2}\\
    \rowcolor{Gray}
    SSAM-Base \cite{yadav2024audiomambaselectivestate} & AS & $69.3$ & 93.2\ci{1.1} & \textbf{97.7\ci{0.1}} & 96.9\ci{0.1} & 70.5\ci{0.5} & 70.3\ci{0.2} & 63.5\ci{0.2} & 87.9\ci{0.3} & 50.4\ci{0.7} & 81.0\ci{0.3} & 52.2\ci{0.1} & 87.7\ci{0.3}\\
    
    \midrule
    \multicolumn{3}{l}{\textbf{MAE Based}} & \multicolumn{5}{l}{}\\
    \rowcolor{Gray}
    AudioMAE \cite{huang2022masked} & AS & $86.0$ & 93.7\ci{0.6} & 89.2\ci{0.2} & 86.6\ci{0.2} & 64.5\ci{0.8} & 68.2\ci{0.2} & 42.2\ci{0.2} & 28.6\ci{1.5} & 29.7\ci{1.0} & 60.6\ci{0.4} & 37.9\ci{0.1} & 63.0\ci{0.3}\\
    MSM-MAE-208 \cite{niizumi2022masked} & AS & ${92.7}$ & 95.7\ci{0.7} & 97.3\ci{0.1} & 97.9\ci{0.1} & 69.1\ci{0.5} & 68.7\ci{0.2} & 63.8\ci{0.5} & 85.7\ci{0.3} & 40.3\ci{0.6} & 78.4\ci{0.6} & 49.5\ci{0.1} & 85.1\ci{0.2}\\
    \rowcolor{Gray}
    MAE-Tiny \cite{yadav2024masked} & AS & ${5.4}$ & 95.6\ci{0.5} & 97.1\ci{0.1} & 97.4\ci{0.1} & 66.4\ci{0.7} & 63.2\ci{0.2} & 64.6\ci{0.3} & 74.3\ci{0.8} & 26.4\ci{0.6} & 70.1\ci{0.5} & 41.6\ci{0.1} & 77.6\ci{0.3}\\
    MAE-Base \cite{yadav2024masked} & AS & ${85.1}$ & 96.2\ci{0.3} & 97.4\ci{0.1} & 98.3\ci{0.1} & 68.3\ci{0.4} & 72.2\ci{0.2} & 67.3\ci{0.3} & 89.4\ci{0.3} & 43.1\ci{0.9} & 80.9\ci{0.4} & 50.4\ci{0.1} & 88.1\ci{0.2} \\
    \rowcolor{Gray}
    % MAE-Large \cite{yadav2024masked} & AS & ${310.0}$ & 95.8\ci{0.6} & 97.5\ci{0.1} & 98.2\ci{0.1} & 69.5\ci{0.6} & 72.4\ci{0.1} & 66.8\ci{0.4} & 90.9\ci{0.2} & 43.6\ci{0.4} & 79.7\ci{0.3} & 50.7\ci{0.1} & 88.5\ci{0.2} \\
    MAE-Huge \cite{yadav2024masked} & AS & ${629.8}$ & \textbf{96.8\ci{0.2}} & 97.5\ci{0.0} & \textbf{98.5\ci{0.0}} & 67.6\ci{0.6} & 71.1\ci{0.2} & 67.1\ci{0.2} & 89.6\ci{0.1} & 40.0\ci{0.7} & 78.3\ci{0.4} & 49.5\ci{0.2} & 86.9\ci{0.1} \\
    MWMAE-Tiny \cite{yadav2024masked} & AS & ${5.4}$ & 93.3\ci{1.0} & 97.1\ci{0.1} & 97.6\ci{0.1} & 68.1\ci{0.4} & 64.4\ci{0.2} & 65.5\ci{0.3} & 77.0\ci{0.6} & 28.6\ci{1.1} & 71.9\ci{0.5} & 43.4\ci{0.1} & 79.1\ci{0.3}\\
    \rowcolor{Gray}
    MWMAE-Base \cite{yadav2024masked} & AS & ${85.1}$ & 96.0\ci{0.5} & 97.4\ci{0.1} & 97.9\ci{0.1} & 69.3\ci{0.6} & 73.1\ci{0.3} & 68.8\ci{0.2} & 90.9\ci{0.2} & 44.2\ci{0.9} & 81.2\ci{0.4} & 51.2\ci{0.2} & 89.2\ci{0.2}\\
    % MWMAE-Large \cite{yadav2024masked} & AS & ${310.0}$ & 95.7\ci{0.5} & 97.4\ci{0.0} & 98.1\ci{0.1} & 70.7\ci{0.6} & 75.5\ci{0.2} & 70.1\ci{0.3} & 93.2\ci{0.1} & 51.9\ci{0.8} & 82.5\ci{0.5} & 53.3\ci{0.1} & 92.6\ci{0.2}\\
    MWMAE-Huge \cite{yadav2024masked} & AS & ${629.8}$ & \textbf{96.8\ci{0.2}} & 97.4\ci{0.0} & 98.2\ci{0.1} & 70.8\ci{0.5} & 74.8\ci{0.1} & \textbf{69.5\ci{0.4}} & 92.4\ci{0.2} & 47.5\ci{0.6} & 81.6\ci{0.4} & 52.1\ci{0.1} & 91.2\ci{0.2} \\
    % \rowcolor{Gray}
    \midrule
    \multicolumn{3}{l}{\textbf{Proposed}} & \multicolumn{5}{l}{}\\
    \rowcolor{Gray}
    AudioMAE++-Tiny & AS & ${8.9}$ & 93.3\ci{0.9} & 96.7\ci{0.1} & 95.1\ci{0.1} & 64.0\ci{1.0} & 65.1\ci{0.2} & 63.1\ci{0.4} & 74.0\ci{0.5} & 32.1\ci{0.8} & 76.6\ci{0.4} & 45.8\ci{0.1} & 78.8\ci{0.3}\\
    AudioMAE++-Base  & AS & ${141.9}$ & 94.3\ci{0.7} & 97.3\ci{0.1} & 97.1\ci{0.1} & 71.2\ci{0.8} & 73.6\ci{0.2} & 65.8\ci{0.4} & 94.0\ci{0.1} & 56.4\ci{0.7} & 85.3\ci{0.4} & 54.0\ci{0.1} & 91.8\ci{0.2}\\
    \rowcolor{Gray}
    AudioMAE++-Large & AS & ${504.0}$ & 94.1\ci{0.4} & 97.4\ci{0.1} & 97.3\ci{0.1} & \textbf{71.7\ci{0.9}} & \textbf{75.1\ci{0.3}} & 66.0\ci{0.3} & \textbf{94.9\ci{0.2}} & 62.1\ci{0.7} & \underline{86.9\ci{0.4}} & \underline{55.3\ci{0.1}} & \textbf{93.7\ci{0.2}}\\
    \bottomrule
    \end{tabular}
    \caption{Comparing AudioMAE++ with existing approaches. LS, AS, VP, LL stand for LibriSpeech, AudioSet, VoxPopuli and LibriLight datasets, respectively. Only encoder parameters of MAE type models are listed.}
    \label{tab:overall}
\end{table*}
\subsection{Implementation details}
\label{ssec:impdetails}
\textbf{Spectrogram features:} A sampling rate of 16,000 Hz is used for all audio clips. We use log-scaled mel spectrogram features with $F=80$ mel-spaced frequency bins as input for our models, extracted with a window size of $25$ ms and a hop size of $10$ ms.\newline
\textbf{Pretraining:} An input spectrogram of shape $[\mathtt{200} \times \mathtt{80}]$ (along time and frequency, respectively), corresponding to a randomly cropped 2-second audio clip is used for pretraining all the proposed models. This is followed by a patch embedding layer that computes rectangular non-overlapping patches of shape $[4\times16]$.
Our default encoder configuration consists of $l=12$ stacked transformer++ blocks with a model feature dimension of $d_m=768$, same as that of a ViT-Base \cite{dosovitskiy2021an} encoder, paired with a decoder comprising of $l=4$ transformer++ blocks with $d_{dec}=384$ dimensions, similar to \cite{niizumi2022masked, yadav2024masked}. 
Additionally, we also evaluate the Tiny ($d_m=192$, $l=12$) and the Large ($d_m=1024$, $l=24$) encoder configurations. The large encoder is paired with a larger decoder, with $d_{dec}=512$ dimensions. In early experiments, we discovered that using RoPE hurts performance (as demonstrated in Sec~\ref{ssec:ablations}), and is thus not used unless specified. 
Similar to previous work \cite{yadav2024masked}, we use an AdamW optimizer with a batch size of 1024 and a weight decay of 0.05 to train all models for 100 epochs, and the learning rate follows a linear warmup (10 epochs) + cosine-decay schedule. Similar to \cite{huang2022masked, yadav2024masked}, no data augmentations were used.\newline
% Similar to previous works on MAE \cite{huang2022masked, yadav2024masked}, we do not use any data augmentations during pretraining.\newline
% Inline with previous works \cite{huang2022masked, yadav2024masked}, no data augmentations were used.\newline
\textbf{Downstream evaluation:} For each downstream task, we extract feature vectors independent of the input audio duration by taking the mean over time across features extracted by models in 2-second audio chunks. We then train an MLP classifier with 1 hidden layer of 1024 neurons for each task, using the official \textit{hear-eval-kit} accompanying the HEAR benchmark. All experiments are repeated with 10 different random seeds, and 95\% confidence intervals are reported. In contrast to the default hyperparameter grid in the \textit{hear-eval-kit}, we use a restricted grid, same as \cite{yadav2024masked}, to aid direct comparison.

\section{Results and ablations}
\label{sec:results}
\subsection{AudioMAE++ versus existing SSL approaches}
Table~\ref{tab:overall} depicts how AudioMAE++ perform in comparison to various recent audio feature representations. For pretrained models from the cited baselines, we extracted fixed feature vectors from available pretrained models and conducted our own downstream experiments, as highlighted in Section~\ref{ssec:impdetails}. 
While we include several recent masked modeling approaches and have included a dedicated section for MAE-based approaches, our main objective is to evaluate the performance of AudioMAE++ with the most directly comparable MAE baselines from \cite{yadav2024masked}: we have identical input clip duration, feature extraction, optimization and downstream evaluation setup with all models from \cite{yadav2024masked}, and compare with parametrically closest configurations. 
The feature embeddings of all the masked autoencoder based approaches are identical for identical encoder configurations, but they can vary across other models. This is in compliance with established frameworks for self-supervised audio representation evaluation, such as HEAR.
% Ideally, one-to-one comparison would need exact feature embedding dimensions, but retraining all models is infeasible, and we are compliant with established frameworks for self-supervised audio representation evaluation, such as HEAR.
All featured masked autoencoder based approaches are pretrained on AudioSet. Finally, we would like to note that we did not include the recent Dasheng \cite{dinkel24b_interspeech} MAE models in our analysis, as they are trained on $52\times$ more data and go beyond a billion parameters,  
% like to note that we did not include Dasheng \cite{dinkel24b_interspeech} models in our analysis; while they are  Dasheng models are trained on much larger audio dataset and go beyond a billion parameters, 
and it was infeasible for us to conduct a comparison at that scale.

% From the table, we can see that the AudioMAE++ models outperform existing MAE based approaches. Our base configuration (AudioMAE++-Base) has the highest overall score of $92.0${\ci{0.2}} amongst all the evaluated models, getting the best results for 2 tasks (NS-5h and SC-5h) across the board, and the best results amongst self-supervised approaches (underlined) for 3 tasks (VL, E50, FSD50K). For Audio tasks, fully supervised PaSST-Base performs the best, but AudioMAE++-Base achieves the best performance amongst the self-supervised models. It's interesting to note that the proposed AudioMAE++-Base outperforms SSL representations specializing on ASR (W2V2, WavLM, HuBERT) on all the speech based tasks, including keyword spotting (SC-5h).
% From the table, we can see that our best model (AudioMAE++-Base) has the highest overall score of $92.0${\ci{0.2}} amongst all the evaluated models, getting the best results for 2 tasks (NS-5h and SC-5h) across the board. For Audio tasks (E50, F50K), fully supervised PaSST-Base performs the best, but AudioMAE++-Base achieves the best performance amongst the self-supervised models. It's interesting to note that AudioMAE++-Base outperforms SSL representations specializing on ASR (W2V2, WavLM, HuBERT) on all the speech based tasks, including keyword spotting (SC-5h) and language identification (VL). 

From the table, we can see that AudioMAE++ Large and Base, with overall scores of $93.7${\ci{0.2}} and $91.8${\ci{0.2}}, respectively, are the top performers amongst all evaluated models, with AudioMAE++-Large getting the best results for 3 tasks (NS-5h, CD, SC-5h) across the board. For Audio tasks (E50, F50K), fully supervised PaSST-Base performs the best, but AudioMAE++-Large and Base are again the top performers amongst the self-supervised models. It is interesting to note that both AudioMAE++ Base and Large outperform SSL representations specializing on ASR (W2V2, WavLM, HuBERT) on emotion recognition (CD) and keyword spotting (SC-5h). 

AudioMAE++ models outperform all vanilla transformer block based MAE approaches by a considerable margin, and also perform well compared to MWMAE models, which possess specialized multihead attention modules to capture local-global attention. 
The fact that AudioMAE++-Base model yields better overall performance than MWMAE-Huge suggests that AudioMAE++ models can potentially capture local-global information despite lacking explicit inductive biases to do so. 
However, it is worth noting that MWMAE models perform better on 3 of 4 music and pitch perception tasks, as well as on speaker count estimation (LC). Finally, while the proposed AudioMAE++ models do have more parameters than directly comparable MAE/MW-MAE counterparts, they show much better scaling characteristics, with AudioMAE++-Base yielding better performance than MAE-Huge and MWMAE-Huge models while possessing $4\times$ fewer parameters. 
Overall, we believe that our empirical analysis demonstrates the viability of the AudioMAE++ approach as well as the underlying transformer++ building block when it comes to learning general-purpose audio representations through self-supervision.

\subsection{Ablation experiments}
\label{ssec:ablations}
\textbf{Impact of RoPE on overall performance:} Table~\ref{tab:rope} shows the impact of using the optional rotary positional embeddings at various staged of the masked autoencoder. We started out with both the encoder and the decoder having RoPE, and then proceed to remove RoPE first from the encoder and then from both encoder and the decoder. It is evident that RoPE does not improve overall performance in the proposed AudioMAE++ framework for the evaluated tasks, as opposed to \cite{hou24_interspeech}, where RoPE improved ASR performance. We believe that this can stem from both the reduced input audio duration (2-sec random crops), which does not require long sequential context modeling abilities, as well as the fact that we evaluate exclusively on utterance level tasks. 
\begin{table}[h]
    % \small
    % \setlength\tabcolsep{10pt}
    \vspace{-1pt}
    \centering
    \begin{tabular}{l|cccc}
    \toprule
    Configuration &  Encoder & Decoder & $s(m)$     \\
    \midrule
    Base & \checkmark & \checkmark  & 91.4\ci{0.2} \\
    Base & $\times$ & \checkmark  & 91.4\ci{0.2} \\
    Base & $\times$ & $\times$  & 92.0\ci{0.2} \\
    \bottomrule
    \end{tabular}
    \caption{Performance impact of RoPE in encoder/decoder.}
    \label{tab:rope}
    % \vspace{-1em}
\end{table}
\newline\textbf{Impact of decoder dimensions $d_{dec}$:} We investigate the impact of decoder complexity on downstream performance by ablating $d_{dec}$ while using the large encoder configuration to ensure that it is not the bottleneck. From Table~\ref{tab:decoder}, we can observe that a $d_{dec}$ of $512$ improves performance considerably over a $384$-dim decoder, whereas increasing it further to $768$ does not further impact performance, highlighting that while the MAE architecture enables an asymmetrical encoder-decoder pairing, decoder complexity can still be a bottleneck when scaling up encoders.
\begin{table}[h]
    \vspace{-1pt}
    \centering
    \begin{tabular}{l|cc}
    \toprule
    Encoder &  $d_{dec}$ & $s(m)$     \\
    \midrule
    Large & 384 & 91.7\ci{0.1} \\
    Large & 512 & \textbf{93.7\ci{0.2}} \\
    Large & 768  & 93.4\ci{0.2} \\
    \bottomrule
    \end{tabular}
    \caption{Decoder complexity and overall performance.}
    \label{tab:decoder}
    \vspace{-10pt}
\end{table}

\section{Conclusion}
We present AudioMAE++, a masked autoencoder for training general-purpose audio representations revamped with transformer++ blocks that are comprised of macaron-style feedforward networks and gated linear units. We compare the proposed AudioMAE++ models with Masked Autoencoders with standard transformer blocks through an extensive empirical analysis on ten varied downstream audio recognition tasks. Pretrained on the AudioSet dataset, AudioMAE++-Base, our best performing model configuration, outperforms all SSL baselines in 5 tasks and give better overall performance than MAE configurations with $4\times$ more parameters, offering much better scaling characteristics. 
Through ablation experiments, we also evaluated the performance impact of rotary positional embeddings (RoPE) within a masked autoencoder framework, and found that RoPE hurts overall downstream performance, in contrast to recent work in the automatic speech recognition domain. A more exhaustive analysis of the interplay of gated linear units, local-global information processing and sequence length would be a good place to start for future research. Overall, based on our analysis, we can conclude that AudioMAE++ is a viable neural architecture for self-supervised audio representation learning.

% To start a new column (but not a new page) and help balance the last-page
% column length use \vfill\pagebreak.
% -------------------------------------------------------------------------
% \vfill\pagebreak

% References should be produced using the bibtex program from suitable
% BiBTeX files (here: strings, refs, manuals). The IEEEbib.bst bibliography
% style file from IEEE produces unsorted bibliography list.
% -------------------------------------------------------------------------

% \newpage
% {\footnotesize
{\ninept
\bibliographystyle{IEEEbib}
\bibliography{strings,refs}

\begin{thebibliography}{10}

\bibitem{bert2019}
Jacob Devlin, Ming-Wei Chang, Kenton Lee, and Kristina Toutanova,
\newblock ``{BERT}: Pre-training of deep bidirectional transformers for language understanding,''
\newblock in {\em Proceedings of the 2019 Conference of the North {A}merican Chapter of the Association for Computational Linguistics: Human Language Technologies, Volume 1 (Long and Short Papers)}, 2019.

\bibitem{xie2022simmim}
Zhenda Xie, Zheng Zhang, Yue Cao, Yutong Lin, Jianmin Bao, Zhuliang Yao, Qi~Dai, and Han Hu,
\newblock ``Simmim: A simple framework for masked image modeling,''
\newblock in {\em Proceedings of the IEEE/CVF Conference on Computer Vision and Pattern Recognition}, 2022, pp. 9653--9663.

\bibitem{bao2022beit}
Hangbo Bao, Li~Dong, Songhao Piao, and Furu Wei,
\newblock ``{BE}it: {BERT} pre-training of image transformers,''
\newblock in {\em International Conference on Learning Representations}, 2022.

\bibitem{baevski2020wav2vec}
Alexei {Baevski}, Yuhao {Zhou}, Abdelrahman {Mohamed}, and Michael {Auli},
\newblock ``wav2vec 2.0: A framework for self-supervised learning of speech representations,''
\newblock in {\em Advances in Neural Information Processing Systems}, 2020, vol.~33, pp. 12449--12460.

\bibitem{chen2022wavlm}
Sanyuan Chen, Chengyi Wang, Zhengyang Chen, Yu~Wu, Shujie Liu, Zhuo Chen, Jinyu Li, Naoyuki Kanda, Takuya Yoshioka, Xiong Xiao, et~al.,
\newblock ``Wavlm: Large-scale self-supervised pre-training for full stack speech processing,''
\newblock {\em IEEE Journal of Selected Topics in Signal Processing}, vol. 16, no. 6, pp. 1505--1518, 2022.

\bibitem{hsu2021hubert}
Wei-Ning {Hsu}, Benjamin {Bolte}, Yao-Hung~Hubert {Tsai}, Kushal {Lakhotia}, Ruslan {Salakhutdinov}, and Abdelrahman {Mohamed},
\newblock ``Hubert: Self-supervised speech representation learning by masked prediction of hidden units.,''
\newblock {\em IEEE Transactions on Audio, Speech, and Language Processing}, pp. 1--1, 2021.

\bibitem{baevski2023efficient}
Alexei Baevski, Arun Babu, Wei-Ning Hsu, and Michael Auli,
\newblock ``Efficient self-supervised learning with contextualized target representations for vision, speech and language,''
\newblock in {\em International conference on machine learning}. PMLR, 2023, pp. 1416--1429.

\bibitem{he2022masked}
Kaiming He, Xinlei Chen, Saining Xie, Yanghao Li, Piotr Doll{\'a}r, and Ross Girshick,
\newblock ``Masked autoencoders are scalable vision learners,''
\newblock in {\em Proceedings of the IEEE/CVF Conference on Computer Vision and Pattern Recognition}, 2022, pp. 16000--16009.

\bibitem{feichtenhofer2022masked}
Christoph Feichtenhofer, Yanghao Li, Kaiming He, et~al.,
\newblock ``Masked autoencoders as spatiotemporal learners,''
\newblock {\em Advances in neural information processing systems}, vol. 35, pp. 35946--35958, 2022.

\bibitem{tong2022videomae}
Zhan Tong, Yibing Song, Jue Wang, and Limin Wang,
\newblock ``Videomae: Masked autoencoders are data-efficient learners for self-supervised video pre-training,''
\newblock {\em Advances in neural information processing systems}, vol. 35, pp. 10078--10093, 2022.

\bibitem{woo2023convnext}
Sanghyun Woo, Shoubhik Debnath, Ronghang Hu, Xinlei Chen, Zhuang Liu, In~So Kweon, and Saining Xie,
\newblock ``Convnext v2: Co-designing and scaling convnets with masked autoencoders,''
\newblock in {\em Proceedings of the IEEE/CVF conference on computer vision and pattern recognition}, 2023, pp. 16133--16142.

\bibitem{scalemae}
Colorado~J Reed, Ritwik Gupta, Shufan Li, Sarah Brockman, Christopher Funk, Brian Clipp, Kurt Keutzer, Salvatore Candido, Matt Uyttendaele, and Trevor Darrell,
\newblock ``Scale-mae: A scale-aware masked autoencoder for multiscale geospatial representation learning,''
\newblock in {\em Proceedings of the IEEE/CVF International Conference on Computer Vision (ICCV)}, October 2023, pp. 4088--4099.

\bibitem{bachmann2022multimae}
Roman Bachmann, David Mizrahi, Andrei Atanov, and Amir Zamir,
\newblock ``Multimae: Multi-modal multi-task masked autoencoders,''
\newblock in {\em European Conference on Computer Vision}. Springer, 2022, pp. 348--367.

\bibitem{zhou2023self}
Lei Zhou, Huidong Liu, Joseph Bae, Junjun He, Dimitris Samaras, and Prateek Prasanna,
\newblock ``Self pre-training with masked autoencoders for medical image classification and segmentation,''
\newblock in {\em 2023 IEEE 20th International Symposium on Biomedical Imaging (ISBI)}. IEEE, 2023, pp. 1--6.

\bibitem{niizumi2022masked}
Daisuke Niizumi, Daiki Takeuchi, Yasunori Ohishi, Noboru Harada, and Kunio Kashino,
\newblock ``Masked spectrogram modeling using masked autoencoders for learning general-purpose audio representation,''
\newblock in {\em HEAR: Holistic Evaluation of Audio Representations}. PMLR, 2022, pp. 1--24.

\bibitem{baade_mae-ast_2022}
Alan Baade, Puyuan Peng, and David Harwath,
\newblock ``{MAE}-{AST}: {Masked} {Autoencoding} {Audio} {Spectrogram} {Transformer},''
\newblock in {\em Interspeech 2022}. Sept. 2022, pp. 2438--2442, ISCA.

\bibitem{huang2022masked}
Po-Yao Huang, Hu~Xu, Juncheng Li, Alexei Baevski, Michael Auli, Wojciech Galuba, Florian Metze, and Christoph Feichtenhofer,
\newblock ``Masked autoencoders that listen,''
\newblock {\em Advances in Neural Information Processing Systems}, vol. 35, pp. 28708--28720, 2022.

\bibitem{yadav2024masked}
Sarthak Yadav, Sergios Theodoridis, Lars~Kai Hansen, and Zheng-Hua Tan,
\newblock ``Masked autoencoders with multi-window local-global attention are better audio learners,''
\newblock in {\em The Twelfth International Conference on Learning Representations}, 2024.

\bibitem{dinkel24b_interspeech}
Heinrich Dinkel, Zhiyong Yan, Yongqing Wang, Junbo Zhang, Yujun Wang, and Bin Wang,
\newblock ``Scaling up masked audio encoder learning for general audio classification,''
\newblock in {\em Interspeech 2024}, 2024, pp. 547--551.

\bibitem{shazeer2020glu}
Noam Shazeer,
\newblock ``Glu variants improve transformer,''
\newblock {\em arXiv preprint arXiv:2002.05202}, 2020.

\bibitem{sun2023length}
Yutao Sun, Li~Dong, Barun Patra, Shuming Ma, Shaohan Huang, Alon Benhaim, Vishrav Chaudhary, Xia Song, and Furu Wei,
\newblock ``A length-extrapolatable transformer,''
\newblock in {\em The 61st Annual Meeting Of The Association For Computational Linguistics}, 2023.

\bibitem{su2024roformer}
Jianlin Su, Murtadha Ahmed, Yu~Lu, Shengfeng Pan, Wen Bo, and Yunfeng Liu,
\newblock ``Roformer: Enhanced transformer with rotary position embedding,''
\newblock {\em Neurocomputing}, vol. 568, pp. 127063, 2024.

\bibitem{lu2020understanding}
Yiping Lu, Zhuohan Li, Di~He, Zhiqing Sun, Bin Dong, Tao Qin, Liwei Wang, and Tie-yan Liu,
\newblock ``Understanding and improving transformer from a multi-particle dynamic system point of view.,''
\newblock in {\em ICLR 2020 Workshop on Integration of Deep Neural Models and Differential Equations}, 2020.

\bibitem{hou24_interspeech}
Zejiang Hou, Goeric Huybrechts, Anshu Bhatia, Daniel Garcia-Romero, Kyu~J. Han, and Katrin Kirchhoff,
\newblock ``Revisiting convolution-free transformer for speech recognition,''
\newblock in {\em Interspeech 2024}, 2024, pp. 4568--4572.

\bibitem{vaswani2017attention}
Ashish Vaswani, Noam Shazeer, Niki Parmar, Jakob Uszkoreit, Llion Jones, Aidan~N Gomez, {\L}ukasz Kaiser, and Illia Polosukhin,
\newblock ``Attention is all you need,''
\newblock {\em Advances in neural information processing systems}, vol. 30, 2017.

\bibitem{transformerxl}
Zihang Dai, Zhilin Yang, Yiming Yang, Jaime Carbonell, Quoc Le, and Ruslan Salakhutdinov,
\newblock ``Transformer-{XL}: Attentive language models beyond a fixed-length context,''
\newblock in {\em Proceedings of the 57th Annual Meeting of the Association for Computational Linguistics}, Florence, Italy, July 2019, pp. 2978--2988, Association for Computational Linguistics.

\bibitem{dauphin2017language}
Yann~N Dauphin, Angela Fan, Michael Auli, and David Grangier,
\newblock ``Language modeling with gated convolutional networks,''
\newblock in {\em International conference on machine learning}. PMLR, 2017, pp. 933--941.

\bibitem{ramachandran2017searching}
Prajit Ramachandran, Barret Zoph, and Quoc~V Le,
\newblock ``Searching for activation functions,''
\newblock {\em arXiv preprint arXiv:1710.05941}, 2017.

\bibitem{dosovitskiy2021an}
Alexey Dosovitskiy, Lucas Beyer, Alexander Kolesnikov, Dirk Weissenborn, Xiaohua Zhai, Thomas Unterthiner, Mostafa Dehghani, Matthias Minderer, Georg Heigold, Sylvain Gelly, Jakob Uszkoreit, and Neil Houlsby,
\newblock ``An image is worth 16x16 words: Transformers for image recognition at scale,''
\newblock in {\em International Conference on Learning Representations}, 2021.

\bibitem{gemmeke2017audio}
Jort~F Gemmeke, Daniel~PW Ellis, Dylan Freedman, Aren Jansen, Wade Lawrence, R~Channing Moore, Manoj Plakal, and Marvin Ritter,
\newblock ``Audio set: An ontology and human-labeled dataset for audio events,''
\newblock in {\em 2017 IEEE international conference on acoustics, speech and signal processing (ICASSP)}. IEEE, 2017, pp. 776--780.

\bibitem{yadav2024audiomambaselectivestate}
Sarthak Yadav and Zheng-Hua Tan,
\newblock ``Audio mamba: Selective state spaces for self-supervised audio representations,''
\newblock in {\em Proc. {INTERSPEECH} 2024 -- 25\textsuperscript{th} Annual Conference of the International Speech Communication Association}, {Kos Island, Greece}, {Sep.} 2024.

\bibitem{koutini22_interspeech}
Khaled Koutini, Jan Schlüter, Hamid Eghbal-zadeh, and Gerhard Widmer,
\newblock ``{Efficient Training of Audio Transformers with Patchout},''
\newblock in {\em Proc. Interspeech 2022}, 2022, pp. 2753--2757.

\bibitem{chenbeats23}
Sanyuan Chen, Yu~Wu, Chengyi Wang, Shujie Liu, Daniel Tompkins, Zhuo Chen, Wanxiang Che, Xiangzhan Yu, and Furu Wei,
\newblock ``Beats: audio pre-training with acoustic tokenizers,''
\newblock in {\em Proceedings of the 40th International Conference on Machine Learning}, 2023, pp. 5178--5193.

\bibitem{gong2022ssast}
Yuan Gong, Cheng-I Lai, Yu-An Chung, and James Glass,
\newblock ``Ssast: Self-supervised audio spectrogram transformer,''
\newblock in {\em Proceedings of the AAAI Conference on Artificial Intelligence}, 2022, vol.~36, pp. 10699--10709.

\end{thebibliography}
}
\end{document}